\documentclass[reqno]{amsart}

\usepackage{ifthen,version}
\newboolean{arxiv-version} % newbooleans are initially false
\setboolean{arxiv-version}{true}
\ifthenelse{\boolean{arxiv-version}}{\includeversion{prnt-arxiv}}%
{\excludeversion{prnt-arxiv}}
\ifthenelse{\boolean{arxiv-version}}{\excludeversion{prnt-cready}}%
{\includeversion{prnt-cready}}

\usepackage{booktabs} % For formal tables
\usepackage{IEEEtrantools}
\usepackage{amssymb,latexsym,amsfonts,amsmath,amsthm}
\usepackage{thmtools, thm-restate}

\usepackage{mdframed,lipsum}

\usepackage{graphicx}

\usepackage{mathrsfs}

\usepackage{graphicx}
\usepackage{paralist}
\usepackage{dsfont}
\usepackage{eso-pic}
\usepackage{epsfig}
\usepackage{mathtools}
\usepackage{epstopdf}

\def\baselinestretch{1}

\usepackage{tikz}
\usetikzlibrary{calc,shapes,arrows}

\usepackage{tikz,circuitikz}
\usetikzlibrary{patterns,calc,angles,quotes,automata,positioning,decorations.markings,arrows,intersections,circuits.logic.US,shapes.arrows,shapes.gates.logic.US,backgrounds}
\usetikzlibrary{automata,positioning,decorations.markings,arrows,intersections,calc,shapes}

\colorlet{darkgreen}{green!40!black}
\colorlet{darkblue}{blue!60!black}
\colorlet{darkred}{red!50!black}
\colorlet{safecellcolor}{yellow!5}
\colorlet{goodcellcolor}{green!10}
\colorlet{badcellcolor}{blue!10}

\tikzset{
	>=stealth,
	box state/.style={draw,rectangle,minimum size=8mm},
	prob state/.style={draw,very thick,shape=circle,darkblue,minimum size=3mm,inner sep=0mm},
	node distance=2cm,on grid,auto, initial text=,
	every loop/.style={shorten >=0pt},
	accepting/.style={double distance=1.2pt, outer sep = 0.6pt+\pgflinewidth},
	accepting dot/.style={above=-2.5pt,circle,fill,darkgreen,inner sep=2pt,radius=1pt},
	loop above/.append style={every loop/.append style={out=120, in=60, looseness=6}},
	loop below/.append style={every loop/.append style={out=300, in=240, looseness=6}},
	loop left/.append style={every loop/.append style={out=210, in=150, looseness=6}},
	loop right/.append style={every loop/.append style={out=30, in=330, looseness=6}},
	accepting arc/.style={dashed},
	marked/.style={
		dashed,
		opacity=0.3
	},
	marked on/.style={alt=#1{marked}{}},
}
\ctikzset{bipoles/not port/width=0.75,bipoles/not port/height=0.5}
\ctikzset{bipoles/not port/circle width=0.3}
\ctikzset{tripoles/american nand port/circle width=0.2}

\usepackage{algorithm}
\usepackage{algorithmic}

\usepackage{xspace}

\newtheorem{theorem}{Theorem}[section]

\newtheorem{problem}[theorem]{Problem}
\newtheorem{proposition}[theorem]{Proposition}

\newtheorem{definition}[theorem]{Definition}
\newtheorem{example}[theorem]{Example}

\newtheorem{remark}[theorem]{Remark}

\numberwithin{equation}{section}

\newcommand{\N}{{\mathbb{N}}}
\newcommand{\Nat}{{\mathbb{N}}}
\newcommand{\Nplus}{{\mathbb{N}}_{> 0}}
\newcommand{\Real}{{\mathbb{R}}}
\newcommand{\Rpos}{{\mathbb{R}}_{> 0}}
\newcommand{\Rplus}{{\mathbb{R}}_{\geq 0}}

\newcommand{\DIST}{{\mathcal D}}

\newcommand{\Mm}{\mathcal{M}}

\newcommand{\AP}{AP} 
\newcommand{\alphabeth}{\mathsf{\Sigma}_{\textsf{a}}}
\newcommand{\word}{\omega}
\newcommand{\Lab}{\mathsf{L}}

\newcommand{\until}{\mathbin{\sf U}}

\newcommand{\nex}{\mathord{\bigcirc}}

\begin{document}

\begin{abstract}
A novel reinforcement learning scheme to synthesize policies for continuous-space Markov decision processes (MDPs) is proposed. 
This scheme enables one to apply model-free, off-the-shelf reinforcement
learning algorithms for finite MDPs to compute optimal strategies for the
corresponding continuous-space MDPs without explicitly constructing the
finite-state abstraction.
The proposed approach is based on abstracting the system with a finite MDP (without constructing it explicitly) with
\emph{unknown} transition probabilities, synthesizing strategies over the abstract
MDP, and then mapping the results back over the concrete continuous-space MDP
with \emph{approximate optimality guarantees}.
The properties of interest for the system belong to a fragment of linear temporal logic,
known as syntactically co-safe linear temporal logic (scLTL), and the synthesis
requirement is to maximize the probability of satisfaction within a given
bounded time horizon.
A key contribution of the paper is to leverage the classical convergence results for reinforcement learning on
finite MDPs and provide control strategies maximizing the probability
of satisfaction over unknown, continuous-space MDPs while providing probabilistic closeness
guarantees.
Automata-based reward functions are often sparse; we present a novel
potential-based reward shaping technique to produce dense rewards to speed up learning. 
The effectiveness of the proposed approach is demonstrated by
applying it to three physical benchmarks concerning the regulation of
a room's temperature, control of a road traffic cell, and of a $7$-dimensional nonlinear model of a BMW $320$i car.
\end{abstract}

\title[Formal Controller Synthesis for Continuous-Space MDPs via Reinforcement Learning]{Formal Controller Synthesis for Continuous-Space MDPs via Model-Free Reinforcement Learning}

\author{Abolfazl Lavaei$^1$}
\author{Fabio Somenzi$^2$}
\author{Sadegh Soudjani$^3$}
\author{Ashutosh Trivedi$^4$}
\author{Majid Zamani$^{4,1}$}
\address{$^1$Department of Computer Science, Ludwig Maximilian University of Munich, Germany.}
\email{lavaei@lmu.de}
\address{$^2$Electrical, Computer \& Energy Engineering, University of Colorado Boulder, USA.}
\email{fabio@colorado.edu}
\address{$^3$School of Computing, Newcastle University, UK.}
\email{sadegh.soudjani@newcastle.ac.uk}
\address{$^4$Department of Computer Science, University of Colorado Boulder, USA.}
\email{\{ashutosh.trivedi,majid.zamani\}@colorado.edu}
\maketitle

\section{Introduction}
{\bf Motivations.} Control systems with stochastic uncertainty can be modeled as Markov decision
processes (MDPs) over uncountable state and action spaces.
These stochastic models have received significant attentions as an important
modeling framework describing many engineering systems; they play significant
roles in many safety-critical applications including power grids and traffic
networks.
Automated controller synthesis~\cite{baier2008principles} for general MDPs to achieve some high-level specifications, e.g., those expressed as linear
temporal logic (LTL) formulae~\cite{pnueli1977temporal}, is inherently
challenging due to its computational complexity and uncountable sets of states and actions.
Closed-form computation of optimal policies for MDPs over uncountable spaces is
not available in general.
One promising approach is to first approximate these models by simpler ones
with finite state sets, perform analysis and synthesis over the abstract
models (using algorithms from formal
methods~\cite{baier2008principles}), and translate the results back over the
original system, while providing guaranteed error bounds in the detour process. 

{\bf Related Literature.} There have been several results, proposed in the past few years, on
abstraction-based synthesis of continuous-space MDPs. Existing results include
construction of finite MDPs for formal verification and synthesis~\cite{APLS08}
and the extension of such techniques to infinite horizon
properties~\cite{tkachev2011infinite} under some strong assumptions over the dynamics. Algorithmic construction of the abstract
models and performing formal synthesis over them are studied in
\cite{SA13,majumdar2019symbolic}. 
Safety verification and formal synthesis of stochastic systems are
respectively studied in~\cite{prajna2007framework} and~\cite{Pushpak2019} using so-called
control barrier certificates. Although the proposed approaches
in~\cite{prajna2007framework} and~\cite{Pushpak2019} do not need the state set
discretization, they require knowing precisely the
probabilistic evolution of states in models which may not be known in general.

Compositional construction of infinite abstractions (reduced-order models) is proposed in~\cite{lavaei2018CDCJ} using dissipativity-type properties of subsystems and their abstractions. Compositional construction of finite MDPs for large-scale stochastic switched systems via $\max$-type small-gain conditions is recently presented in~\cite{lavaei2019HSCC_J}. Compositional construction of finite abstractions is studied in~\cite{lavaei2018ADHSJJ,lavaei2017HSCC} using respectively small-gain and dissipativity-type conditions both for discrete-time stochastic control systems. Compositional construction of finite MDPs for networks of not necessarily stabilizable stochastic systems via relaxed dissipativity conditions is discussed in~\cite{lavaei2019NAHS}. 

To construct finite MDPs of continuous-space ones with guaranteed error bounds between them, we need to establish some sort of similarity relations between them. Similarity relations over finite-state stochastic systems have been studied,
either via exact notions of probabilistic (bi)simulation
relations~\cite{larsen1991bisimulation,segala1995probabilistic} or
approximate versions~\cite{desharnais2008approximate}.
Similarity relations for models with general, uncountable state spaces have also
been proposed in the literature. These relations either depend on stability
requirements on model outputs via martingale theory or the contractivity
analysis~\cite{julius2009approximations} or enforce
structural abstractions of models by exploiting
continuity conditions on their probability laws~\cite{abate2014probabilistic}. A
new bisimilarity relation is proposed in~\cite{SIAM17,HS18_robust} based on the
joint probability distribution of two models and enables combining model
reduction together with space discretization. 

Unfortunately, construction of finite MDPs, studied in the aforementioned
literatures, suffers severely from the so-called \emph{curse of dimensionality}:
the computational complexity of constructing finite MDPs grows
exponentially as the number of state variables increases.
In addition, one needs to know precise models of continuous-space MDPs to
construct those finite abstractions and, hence, the proposed approaches in the relevant literature are not applicable in the settings where the transition structure is unknown. These challenges motivated us to employ reinforcement learning for the
controller synthesis of such complex systems.

Reinforcement learning (RL)~\cite{Sutton18} is an approach to sequential
decision making in which agents rely on reward signals to choose actions aimed
at achieving prescribed objectives.
Model-free reinforcement learning~\cite{Strehl06} refers to techniques that are
asymptotically space-efficient because they do not store the probabilistic
transition structure of the environment. These techniques include algorithms like
TD($\lambda$)~\cite{Sutton1988} and Q-learning~\cite{watkins1989learning} as
well as their extensions to deep neural networks such as deep deterministic
policy gradient (DDPG)~\cite{Lillicrap15} and neural-fitted Q-iterations~\cite{Riedmi05}.  
Model-free reinforcement learning has achieved performance comparable to that of
human experts in video and board games~\cite{Tesaur95,Mnih15,Silver16}.
This success has motivated extensions of reinforcement learning to the control of safety-critical systems~\cite{Lillicrap15,Levine16} in spite of a lack of theoretical
convergence guarantees of reinforcement learning for general continuous state spaces~\cite{dai2017sbeed}.

\textbf{Main contribution.} By utilizing a closeness guarantee between probabilities of
satisfaction by the unknown continuous-space MDP and by its finite abstraction which can be controlled a-priori,
and leveraging the classical convergence results for reinforcement learning on finite-state MDPs, we provide, for the first time, a reinforcement
learning approach for MDPs with uncountable state sets while providing convergence guarantees. In particular, this approach enables us to apply model-free, off-the-shelf reinforcement
learning algorithms to compute $\varepsilon$-optimal strategies for
continuous-space MDPs with a precision $\varepsilon$ that is defined a-priori
and without explicitly constructing finite abstractions.
Another key contribution of the paper is a novel potential-based
reward shaping \cite{Ng99}
technique to produce dense rewards that is based on the structure of
the property automaton.
Although the techniques presented in this paper can be adapted to
model-based RL, the experiments presented in this work deal with
model-free RL.

{\bf Recent Works.} A model-free reinforcement learning framework for
synthesizing policies for unknown, and possibly continuous-state, MDPs is recently
presented
in~\cite{hasanbeig2019certified,hasanbeig2019reinforcement,yuan2019modular}. The proposed approaches
in~\cite{hasanbeig2019certified,hasanbeig2019reinforcement,yuan2019modular}
provide theoretical guarantees only when the MDP has the finite number of states,
and the corresponding results for continuous-state MDPs are empirically
illustrated. The results in \cite{munos} provide a reinforcement learning approach for \emph{deterministic} continuous-space control systems where the closeness between finite approximations and concrete models are only guaranteed asymptotically, rather than according to some formal relations that are in the end required to ensure the correspondence of controllers for temporal logic specifications over model trajectories. The results in \cite{silver14} provide deterministic policy
gradient algorithms for MDPs with continuous state and action spaces using reinforcement learning, but without providing any quantitative guarantee on the optimality of synthesized policies for original MDPs. In contrast, we utilize here a closeness
guarantee between probabilities of satisfaction by the unknown
continuous-space MDP and by its finite abstraction to compute
$\varepsilon$-optimal strategies for original systems using the reinforcement learning with a-priori defined precision $\varepsilon$.

{\bf Organization.} The rest of this paper is organized as follows.  In the next section
we introduce background definitions and notations. Then we formulate
the main problem in the section afterward. In that section, in
particular, we propose closeness guarantees between probabilities of satisfaction by continuous-space MDPs and their
finite-state counterparts and its connection to the classical
convergence results for reinforcement learning on finite-state
MDPs.  Finally, to demonstrate the effectiveness of the proposed
results, we apply our approaches to several physical benchmarks in the last section.

\section{Preliminaries}
As usual, we write $\Nat$ and $\Nplus$ for sets of nonnegative and positive
integers.
Similarly, we write $\Real$, $\Rpos$, and $\Rplus$ for sets of reals,
positive and nonnegative reals, respectively.
For a set of $N$ vectors, $x_1 \in \mathbb R^{n_1},\ldots,x_N \in \mathbb
R^{n_N}$, we write $[x_1;\ldots;x_N]$ to denote the corresponding vector of
dimension $\sum_i n_i$.
Given a vector $x\in\mathbb{R}^{n}$ we write $\Vert x\Vert$ for its Euclidean
norm and for $a\in\mathbb R$ we write $\vert a\vert$ for its
absolute value.

A \emph{discrete probability distribution}, or just distribution, over a
(possibly uncountable) set $X$ is a function $d : X {\to} [0, 1]$ such that 
$\sum_{x \in X} d(x) = 1$ and $\mathit{supp}(d) = \{x \in X \mid  d(x)
{>} 0\}$ is at most countable. 
Let $\DIST(X)$ denote the set of all discrete distributions over $X$.
We consider a probability space $(\Omega,\mathcal F_{\Omega},\mathds{P}_{\Omega})$,
where $\Omega$ is the sample space, $\mathcal F_{\Omega}$ is a sigma-algebra on
$\Omega$ comprising subsets of $\Omega$ as events, and $\mathds{P}_{\Omega}$ is
a probability measure that assigns probabilities to events.
We assume that random variables introduced in this article are measurable
functions of the form
$X:(\Omega,\mathcal F_{\Omega})\rightarrow (S_X,\mathcal F_X)$. 
Any random variable $X$ induces a probability measure on  its space
$(S_X,\mathcal F_X)$ as $Prob\{A\} = \mathds{P}_{\Omega}\{X^{-1}(A)\}$ for any
$A\in \mathcal F_X$. 
When clear from the context, we use the probability measure on $(S_X,\mathcal F_X)$
without mentioning the probability space and the function $X$.

A topological space $S$ is called a {\it Borel space} if it is homeomorphic to a Borel
subset of a Polish space (i.e., a separable and completely metrizable
space).
Examples of a Borel space are Euclidean space $\mathbb R^n$, its Borel
subsets endowed with a subspace topology, as well as hybrid spaces.  
Any Borel space $S$ is assumed to be endowed with a Borel sigma-algebra, which is
denoted by $\mathcal B(S)$. We say that a map $f : S\rightarrow Y$ is measurable
whenever it is Borel measurable.

\subsection{Discrete-Time Stochastic Control Systems}

\begin{definition}
	A {\it discrete-time stochastic control system} (dt-SCS) is a tuple 
	\begin{equation}
	\label{eq:dt-SCS}
	\Sigma=\left(X,U,\varsigma,f\right)\!,
	\end{equation}
	where 
	\begin{itemize}
		\item
		$X\subseteq \mathbb R^n$ is a Borel space as the state space of the
		system. We denote by $(X, \mathcal B (X))$ the measurable space with
		$\mathcal B (X)$  being the Borel sigma-algebra $X$;
		\item
		$U$ is the input space of the system; 
		\item
		$\varsigma$ is a sequence of independent and identically distributed
		(i.i.d.) random variables from a sample space $\Omega$ to the set
		$V_\varsigma$, namely	 
		$\varsigma:=\{\varsigma(k):\Omega\rightarrow V_{\varsigma},\,\,k\in\N\}$;
		\item
		$f:X{\times} U{\times} V_{\varsigma} \rightarrow X$ is a measurable function
		characterizing the state evolution of $\Sigma$.
	\end{itemize}
\end{definition} 
The evolution of the state of dt-SCS $\Sigma$ for an initial state $x(0)\in
X$ and an input sequence $\{\nu(k):\Omega\rightarrow U,\,\,k\in\mathbb N\}$ is
described as: 
\begin{equation}
x(k+1)  = f(x(k),\nu(k),\varsigma(k)).\label{Eq_1a}
\end{equation}

\begin{remark}
	The input space $U$ of a dt-SCS $\Sigma$ is in general a continuous
	Borel space, e.g., a subset of $\mathbb R^m$. Since any input
	sequence will be implemented by a digital controller, w.l.o.g.\ we
	assume that the input space $U$ is finite.
\end{remark}

We define \emph{Markov policies} to control the system in~\eqref{eq:dt-SCS}. 
\begin{definition}
	For the dt-SCS $\Sigma$ in~\eqref{eq:dt-SCS}, a Markov policy is a sequence
	$\rho = (\rho_0,\rho_1,\rho_2,\ldots)$ of universally measurable stochastic kernels $\rho_n$~\cite{Bertsekas1996}, each defined on the input space $U$ given $X$ such that for all $x_n\in X$, $\rho_n(U\,\big|\,x_n)=1$.
	The class of all such Markov policies is denoted by $\bar\Pi_M$. 
\end{definition} 

We associate to $U$ the set $\mathcal U$ to be the collection of sequences $\{\nu(k):\Omega\rightarrow U,\,\,k\in\mathbb N\}$, in which $\nu(k)$ is independent of $\varsigma(t)$ for any $k,t\in\mathbb N$ and $t\ge k$. The random sequence $x_{a\nu}:\Omega \times\mathbb N \rightarrow X$ satisfying~\eqref{Eq_1a} for any initial state $a\in X$, and $\nu(\cdot)\in\mathcal{U}$ is called the \textit{solution process} of $\Sigma$ under the input $\nu$ and the initial state $a$.

\subsection{Requirement Specification in scLTL}\label{property}
Formal requirements provide the rigorous and unambiguous formalism to express
requirements over MDPs~\cite{baier2008principles}.
A common way to describe  formal requirements is by using automata-based
specifications or logic-based specifications using formulae in, for instance, linear temporal logic (LTL).
For example, consider a dt-SCS $\Sigma$ in~\eqref{eq:dt-SCS} and a measurable target set
$\mathsf B\subset X$.
We say that a state trajectory $\{x(k)\}_{k\ge 0}$ reaches a target set
$\mathsf B$ within time interval $[0,T]\subset \mathbb N$, if there exists a
$k\in [0,T]$ such that $x(k)\in \mathsf B$. 
This bounded reaching of $\mathsf B$ is denoted  by $ \lozenge^{\le T}\{x\in
\mathsf B\}$ or briefly $\lozenge^{\le T}\mathsf B$.  For $T\rightarrow \infty$,
we denote  the reachability property as $ \lozenge \mathsf B$, i.e., eventually
$\mathsf B$. 
For a dt-SCS $\Sigma$ with a policy $\rho$,  we want to compute the probability
that a state trajectory reaches $\mathsf B$ within the time horizon $T\in
{\mathbb N}$, i.e.,  
$\mathbb P(\lozenge^{\le T} \mathsf B)$. 
The \emph{reachability probability} is the probability that the target set
$\mathsf B$ is eventually reached and is denoted by  
$\mathbb P(\lozenge \mathsf B)$.
In this paper, we deal with properties more complex than simple reachability
property. 

\noindent{\textbf{Finite Automata}.} A {\it deterministic finite automaton}
(DFA) is a tuple
$\mathcal A = (Q, \mathsf{\Sigma}_{\textsf{a}}, \mathsf{t}, q_0,
F_{\textsf{a}})$ where $Q$ is a finite set of states,
$\mathsf{\Sigma}_{\textsf{a}}$ is an alphabet, 
$\mathsf{t}: Q \times \mathsf{\Sigma}_{\textsf{a}} \to Q$ is a transition function, $q_0 \in Q$ is the
initial state, and  $F_{\textsf{a}} \subseteq Q$ are accepting states.
We write $\lambda$ for the empty string and $\mathsf{\Sigma}_{\textsf{a}}^*$ for
the set of all strings over $\mathsf{\Sigma}_{\textsf{a}}$.
The extended transition function $\mathsf{\hat t}: Q \times
\mathsf{\Sigma}_{\textsf{a}}^* \to Q$ (transition function extended to summarize
the effect of reading a string) can be defined as:
\[
\mathsf{\hat t}(q, \bar w) =
\begin{cases}
q, \!&\quad \text{ if $\bar w = \lambda$},\\
\mathsf{t}(\mathsf{\hat t}(q, x), a), \!& \quad\text{ if $\bar w = xa$ for $x\in\mathsf{\Sigma}_{\textsf{a}}^*$ and \!$a \in \mathsf{\Sigma}_{\textsf{a}}$.} 
\end{cases}
\]

The language $\mathcal{L}(\mathcal A)$ accepted by a DFA $\mathcal A$ is defined as
$\mathcal{L}(\mathcal A) = \{\bar w \::\:   \mathsf{\hat t}(q_0, \bar w) \in F_{\textsf{a}}\}$.
DFAs are well-established models to express regular specifications over finite
words.
DFAs can also be interpreted over $\omega$-words:
an $\omega$-word is accepted if there is a prefix that is accepted by the DFA.
Among others, DFAs are expressive enough to capture syntactically a co-safe
fragment of linear temporal logic (LTL) defined next.

\noindent{\textbf{Linear Temporal Logic}.} Consider a set of atomic propositions
$\AP$ and the alphabet $\alphabeth := 2^{\AP}$.  
Let $\word=\word(0),\word(1),\word(2),\ldots \in\alphabeth^{\mathbb{N}}$ be an
infinite word, that is, a string composed of letters from $\alphabeth$. 
We are interested in those atomic propositions that are relevant to the dt-SCS
via a measurable labeling function $\Lab$ from the state space to the alphabet 
as $\Lab:X\rightarrow \alphabeth$.
State trajectories  $\{x(k)\}_{k\geq 0}\in X^{\mathbb N} $ can be readily
mapped to the set of infinite words $\alphabeth^{\mathbb N}$, as
$\word=\Lab(\{x(k)\}_{k\geq0}):=\{\word\in \alphabeth^{\mathbb N}\,|\,\word(k)=
\Lab(x(k)) \}$.
%\]
Consider LTL properties with the syntax \cite{baier2008principles}\vspace{-0.2cm}
\begin{equation*}
\phi ::=  \operatorname{true} \,|\, p \,|\, \neg \phi \,|\,\phi_1 \wedge \phi_2 \,|\, \nex \phi \,|\, \phi_1\until \phi_2.
\end{equation*}
Let $\word_k=\word(k),\word(k+1),\word(k+2),\ldots  $ be a subsequence (suffix) of $\word$, then  
the satisfaction relation between $\word$ and a property $\phi$, expressed in LTL,  is denoted by $\word\vDash\phi$  
(or equivalently $\word_0\vDash\phi$). 
Semantics of the satisfaction relation are defined recursively over $\word_k$ and
the syntax of the LTL formula $\phi$.
An atomic proposition $ p\in \AP$ is satisfied by $\word_k$, i.e.,  $\word_k\vDash p$, iff   $p \in\word(k)$.  Furthermore, 
$\word_k\vDash \neg \phi$  if $\word_k\nvDash\phi$ and 
we say that  $\word_k\vDash \phi_1\wedge\phi_2$ 
if $ \word_k\vDash \phi_1$ and $\word_k\vDash \phi_2$.
The next operator $\word_k\vDash\nex\phi $ holds if the property holds at the next time instance   $ \word_{k+1}\vDash \phi$. We denote by $\nex^j$, $j\in\N$, $j$ times composition of the next operator. With a slight abuse of the notation, one has $\nex^0\phi=\phi$ for any property $\phi$.
The temporal until operator $\word_k\vDash \phi_1\until\phi_2$  holds if $ \exists i \in \mathbb{N}:$ $\word_{k+i} \vDash \phi_2, \mbox{and } 
\forall j \in{\mathbb{N}:} 0\leq j<i, \word_{k+j}\vDash \phi_1
$.
Based on these semantics, disjunction ($\vee$) can be defined by
$ \word_k\vDash \phi_1\vee\phi_2\ \Leftrightarrow  \  \word_k\vDash \neg(\neg\phi_1 \wedge \neg\phi_2)$.
This paper focuses on a fragment of LTL properties known as syntactically
co-safe linear temporal logic (scLTL) \cite{KupfermanVardi2001} defined
below. 
\begin{definition}[Syntactically Co-Safe LTL (scLTL)]\label{def:scLTL}
	An scLTL over a set of atomic propositions $\AP$ is a fragment of LTL such
	that the negation operator ($\neg$) only occurs before atomic propositions
	characterized by the following grammar:
	\begin{equation*}
	\phi ::=  p \,|\, \neg p \,|\, \phi_1 \lor \phi_2  \,|\,\phi_1 \wedge \phi_2 \,|\, \nex \phi \,|\, \phi_1\until \phi_2.
	\end{equation*}
\end{definition}

Even though scLTL formulas are defined over infinite words (as in LTL formulae),
their satisfaction is guaranteed in the finite time~\cite{KupfermanVardi2001}. Any
infinite word $\word\in\alphabeth^{\mathbb{N}}$ satisfying an scLTL formula
$\phi$ has a finite word $\word_f\in\alphabeth^n$, $n\in\mathbb N$, as its
prefix such that all infinite words with a prefix $\word_f$ also satisfy the
formula $\phi$. We denote the set of all such finite prefixes associated with an
scLTL formula $\phi$ by $\mathcal L_f(\phi)$.

For verification and synthesis purposes, the scLTL properties can be compiled
into a DFA $\mathcal A_{\phi}$ over the alphabet $2^{AP}$ such that $\mathcal L_f(\phi) =
\mathcal L(\mathcal A_{\phi})$~\cite{KupfermanVardi2001}.
This construction is routine; we refer the interested reader to \cite{KupfermanVardi2001} for details of
the construction of the DFA $\mathcal A_\phi$ from $\phi$ such that $\mathcal L(\mathcal
A_{\phi})=\mathcal L_f(\phi)$. 
The resulting DFA has the property that there is a unique accepting state and
all out-going transitions from that state are self-loops.
Such a DFA is also known as a {\it co-safety automaton}.
In the rest of the paper we assume that the DFA $\mathcal{A}_\phi$ for an 
scLTL property $\phi$ is a co-safety automaton.

Given a policy $\rho$, the probability that a state trajectory
of $\Sigma$ satisfies an scLTL property $\phi$ over the time horizon $[0,T]$,
is denoted by $\mathbb P(\word_f \in\mathcal L(\mathcal A_\phi)~\text{s.t.}~
|\word_f|\le T+1)$, where $|\word_f|$ is the length of $\word_f$ \cite{DLT08}.
The co-safety automaton for the bounded time-horizon satisfaction can be
computed by unrolling the DFA for $\phi$.
In the rest of the paper we assume such a representation for the finite horizon satisfaction.

\begin{remark}
	We emphasize that there is no \emph{closed-form solution} for computing
	optimal policies enforcing scLTL specifications over
	\emph{continuous-space} MDPs. One can employ the approximation
	approaches, discussed later, to synthesize those policies which,
	however, suffer severely from the curse of dimensionality and require
	knowing precisely the probabilistic evolution of states in
	models. Instead, we propose in this paper, for the first time, an RL
	approach providing policies for unknown, continuous-space MDPs while
	providing \emph{quantitative guarantees} on the satisfaction of
	properties. 
\end{remark}

\section{Controller Synthesis for Unknown Continuous-Space MDPs}\label{key-results}
We are interested in automatically synthesizing controllers for unknown
continuous-space MDPs whose requirements are provided as scLTL
specifications.
Given a discrete-time stochastic control system
$\Sigma=(X,U,\varsigma,f)$, where $f$ and the distribution of
$\varsigma$ are unknown, and given an scLTL formula $\phi$, we wish to
synthesize a Markov policy enforcing the property $\phi$ over $\Sigma$
with the probability of satisfaction within a guaranteed threshold from
the unknown optimal probability.

In order to provide any formal guarantee, we need to make further assumptions
on the dt-SCS.
In particular, we assume that the dynamical system in \eqref{Eq_1a} is
Lipschitz-continuous with a constant $\mathscr{H}$.
Consider the dynamical system in \eqref{Eq_1a} where
$\varsigma(\cdot)$ is i.i.d.\ with a known distribution $t_\varsigma(\cdot)$. 
Suppose that the vector field $f$ is continuously differentiable and the matrix $\frac{\partial f}{\partial \varsigma}$ is invertible. 
Then, the \emph{implicit function theorem} guarantees the existence and uniqueness of a function $g:X\times X\times U\rightarrow V_\varsigma$ such that $\varsigma(k) = g(x(k+1),x(k),\nu(k))$. 
In this case, the conditional density function is:
\begin{equation*}
t_x(x' | x,\nu) = \left|\det\left[\frac{\partial g}{\partial x'}(x',x,\nu)\right]\right|t_\varsigma(g( x',x,\nu)).
\end{equation*}
The Lipschitz constant $\mathscr{H}$ is specified by the dependency of the function $g(x',x,\nu)$ on the variable $x$.
As a special case consider a nonlinear system with an additive noise
$$f(x,\nu,\varsigma) = f_a(x,\nu)+\varsigma.$$
Then the invertibility of $\frac{\partial f}{\partial \varsigma}$ is guaranteed and $g( x',x,\nu) = x'-f_a(x,\nu)$. In this case,
$\mathscr{H}$ is the product of the Lipschitz constant of $t_\varsigma(\cdot)$ and $f_a(\cdot)$.

The next example provides a systematic way of computing $\mathscr{H}$ for the class of linear continuous-space MDPs with an additive noise.  
\begin{example}
	Consider a dt-SCS $\Sigma$ with linear dynamics $x(k+1) = A
	x(k)+B\nu(k)+\varsigma(k)$, where $A= [a_{ij}]$ and $\varsigma(k)$ are
	i.i.d.\ for $k=0,1,2,\ldots$ with a normal distribution having zero mean
	and covariance matrix $diag$\footnote{$diag(\sigma_1,\ldots,\sigma_n)$
		is a diagonal matrix with $\sigma_1,\ldots,\sigma_n$ as its entries.}$(\sigma_1,\ldots,\sigma_n)$. Then, one obtains $\mathscr{H} = \sum_{i,j}\dfrac{2|a_{ij}|}{\sigma_i\sqrt{2\pi}}$. Note that for the computation of the approximation error (cf. \eqref{eq:metric_lit}), it is sufficient to know an upper bound on entries of the matrix $A$ and a lower bound on the standard deviation of the noise.
\end{example}

An alternative way of computing the Lipschitz constant $\mathscr{H}$ is to
estimate it from sample trajectories of $\Sigma$. This can be done by first
constructing a non-parametric estimation of the conditional density function
using techniques proposed in~\cite{scott:1992} and then compute the Lipschitz constant
numerically using the derivative of the estimated conditional density function. 

More specifically, we can use a conditional kernel density estimation (CKDE) that puts a kernel around each data point. The main purpose of using kernels is to interpolate between the observed data in order to predict the density at the unobserved data points. CKDE provides the following estimation for the conditional density function:
\begin{equation}
\label{eq:estimate}
t_x^{est} (x'|x)= \frac{\sum_{i=1}^{N_{\mathsf s}} K_{\bar h_1}(x'-x_i')K_{\bar h_2}(\|x-x_i\|)}{\sum_{i=1}^{N_{\mathsf s}} K_{\bar h_2}(\|x-x_i\|)},
\end{equation}
where data pairs $(x_i,x_i')$ are extracted from sample trajectories with $x_i'$ being the observed next state for the current state $x_i$, $K_{\bar h}(y) := \frac{1}{\bar h^n}K(\frac{y}{\bar h})$, $K$ is a kernel function, i.e., a symmetric probability distribution with a bounded variance (e.g. the Gaussian), $n$ is the dimension of $x$, and $\bar h$ is the bandwidth controlling the kernel widths. 
This form is known as the Nadaraya-Watson conditional density estimator, which is consistent when $\bar h_1\rightarrow 0$, $\bar h_2\rightarrow 0$, and $N_{\mathsf s}\bar h_1\bar h_2\rightarrow 0$ as $N_{\mathsf s}\rightarrow 0$ \cite{2012arXiv1206.5278H}.
In our case, we can use \eqref{eq:estimate} while making both sides also dependent on the input $\nu$, and then compute its Lipschitz constant numerically.
The numerical computation involves taking the derivative w.r.t.\ $x$, then its norm, and finally maximizing over both $x,x'$. Note that we do not need the best possible Lipschitz constant: any upper bound is also sufficient but at the cost of making the formulated errors more conservative (cf.  errors \eqref{eq:metric_lit}-\eqref{eq:metric_lit2} in Theorem~\ref{thm:key-thm}).

Now we have all required ingredients to state the main problem we address in this paper.

\begin{mdframed}
	\begin{problem}\label{problem}
		Let $\phi$ be an scLTL formula and $\Sigma=(X,U,\varsigma,f)$ a
		continuous-space MDP, where $f$ and the distribution of $\varsigma$ are unknown,
		but the Lipschitz constant $\mathscr{H}$ is known.
		Synthesize a Markov policy that satisfies the property $\phi$ over $\Sigma$ with
		probability within a-priori defined threshold $\varepsilon$ from the unknown
		optimal probability.
	\end{problem}
\end{mdframed}

To present our solution to this problem, we first present a technical result
connecting continuous-space MDPs with corresponding finite MDP abstractions.
We then exploit this result to provide a
reinforcement learning-based solution to Problem~\ref{problem}. We emphasize again that we do not construct explicitly finite abstractions of continuous-space MDPs in this work. In fact, we cannot construct them because the dynamics of continuous-space MDPs are unknown.

\subsection{Abstraction of dt-SCS $\Sigma$ by a Finite MDP}\label{algo:MC_app}

A dt-SCS $\Sigma$ in \eqref{eq:dt-SCS} can be \emph{equivalently} represented as a Markov decision process (MDP) \cite[Proposition 7.6]{kallenberg1997foundations}
\begin{equation}\notag
\Sigma=\left(X,U,T_{\mathsf x}\right)\!,	
\end{equation}
where the map $T_{\mathsf x}:\mathcal B(X)\times X\times U\rightarrow[0,1]$,
is a conditional stochastic kernel that assigns to any $x \in X$, and $\nu\in U$, a probability measure $T_{\mathsf x}(\cdot | x,\nu)$
on the measurable space
$(X,\mathcal B(X))$
so that for any set $\mathcal{A} \in \mathcal B(X)$, 
$$\mathbb P (x(k+1)\in \mathcal{A}\,\big|\, x(k),\nu(k)) = \int_\mathcal{A} T_{\mathsf x} (dx'\,\big|\,x(k),\nu(k)).$$
For given input $\nu(\cdot),$ the stochastic kernel $T_{\mathsf x}$ captures the evolution of the state of $\Sigma$ and can be uniquely determined by the pair $(\varsigma,f)$ from \eqref{eq:dt-SCS}. In other words, $T_{\mathsf x}$ contains the information of the function $f$ and the distribution of the noise $\varsigma(\cdot)$ in the dynamical representation.

Now we approximate a dt-SCS $\Sigma$ with a \emph{finite} $\widehat\Sigma$ using an abstraction algorithm. The algorithm first constructs a
finite partition of the state space $X = \cup_i \mathsf X_i$.
Then representative points $ \bar x_i\in \mathsf X_i$ are selected as abstract states.
Given a dt-SCS $\Sigma=\left(X,U,\varsigma,f\right)$,
the constructed finite MDP $\widehat\Sigma$ is
\begin{equation}
\label{eq:abs_tuple}
\widehat\Sigma =(\hat X,\hat U,\varsigma,\hat f),
\end{equation}
where $\hat X = \{\bar x_i,i=1,\ldots,n_x\}$, a finite subset of $X$, and $\hat U := U$ are finite state and input sets of the MDP~$\widehat\Sigma$. Moreover, $\hat f:\hat X\times\hat U\times V_\varsigma\rightarrow\hat X$ is defined as
$\hat f(\hat{x},\hat{\nu},\varsigma) = \Pi_x(f(\hat{x},\hat{\nu},\varsigma))$,	
where $\Pi_x:X\rightarrow \hat X$ is the map that assigns to any $x\in X$, the representative point $\bar x\in\hat X$ of the corresponding partition set containing $x$.
The initial state of $\widehat\Sigma$ is also selected according to
$\hat x_0 := \Pi_x(x_0)$ with $x_0$ being the initial state of
$\Sigma$.

The proposed dynamical representation employs the map $\Pi_x:X\rightarrow \hat
X$ that assigns to any $x\in X$, the representative point $\bar x\in\hat X$ of
the corresponding partition set containing $x$ satisfying the inequality:
\begin{equation}
\label{eq:Pi_delta}
\Vert \Pi_x(x)-x\Vert \leq \delta,\quad \forall x\in X,
\end{equation}
where $\delta:=\sup\{\|x-x'\|,\,\, x,x'\in \mathsf X_i,\,i=1,2,\ldots,n_x\}$ is the \emph{state} discretization parameter.

Note that one can write the equivalent finite-MDP representation of $\widehat \Sigma$ in~\eqref{eq:abs_tuple} as \cite[Chapter 3.5]{puterman2014markov}
\begin{equation}\label{MDP representation}
\widehat\Sigma =(\hat X,\hat U,\hat T),
\end{equation}
where 
\begin{equation*}
\hat T (x'|x,\nu) 
= T_{\mathsf x} (\Xi(x')|x,\nu), \quad \forall x,x'\in \hat X, \nu\in \hat U,
\end{equation*}
and $\Xi:X\rightarrow 2^X$ is a map that assigns to any $x\in X$, the corresponding partition set it belongs to, i.e.,
$\Xi(x) = \mathsf X_i$ if $x\in \mathsf X_i$ for some $i=1,2,\ldots,n_x$. We employ this finite-MDP representation of \eqref{MDP representation} in Section~\ref{sec:rl}.

The following theorem~\cite{SA13} shows the closeness between a continuous-space MDP $\Sigma$ and its finite abstraction $\widehat\Sigma$ in a probabilistic setting.

\begin{theorem}
	\label{thm:key-thm}
	Let $\Sigma=(X,U,\varsigma,f)$ be a continuous-space MDP  and $\widehat\Sigma=(\hat X, \hat U,\varsigma,\hat f)$ be its finite abstraction. For a given scLTL specification $\phi$, and for any policy $\hat\nu(\cdot)\in\mathcal{\hat U}$ that preserves Markov property for the closed-loop $\widehat\Sigma$ (denoted by $\widehat\Sigma_{\hat \nu}$), the closeness between two systems can be acquired as
	\begin{equation}\label{eq:metric_lit}
	|\mathbb P(\Sigma_{\hat \nu}\vDash\phi) - \mathbb P(\widehat\Sigma_{\hat \nu}\vDash\phi)|\le \varepsilon,\quad \text{with} ~\varepsilon := T \delta \mathscr{H}\mathscr{L}\!,
	\end{equation}
	where $T$ is the finite time horizon, $\delta$ is the state discretization parameter, $\mathscr{H}$ is the Lipschitz constant of the stochastic kernel, and $\mathscr{L}$ is the Lebesgue measure of the specification set.
	Moreover, \emph{optimal probabilities} of satisfying the specification over the two models are different with a distance of at most $2\varepsilon$:
	\begin{equation}
	\label{eq:metric_lit2}
	\big|\max_{\nu\in \bar\Pi_M}\mathbb P(\Sigma_{\nu}\vDash\phi) - \max_{\hat \nu\in \hat{\bar{\Pi}}_M}\mathbb P(\widehat\Sigma_{\hat \nu}\vDash\phi)\big|\le 2\varepsilon,
	\end{equation}
	where $\bar\Pi_M$ and $\hat{\bar{\Pi}}_M$ are sets of Markov policies over $\Sigma$ and $\widehat\Sigma$, respectively.
\end{theorem}

The error bound $\varepsilon$ in \eqref{eq:metric_lit} is obtained by characterizing $\mathbb P(\Sigma_{\hat \nu}\vDash\phi)$ recursively similar to dynamic programs (DP). This error is related to the approximation of the continuous kernel with a discrete one, hence the term $\delta\mathscr{H}$. There is also an integration over the specification set, thus $\mathscr{L}$ appears in $\varepsilon$. Finally, the errors contributed in every iteration of the DP are added, hence the horizon $T$.

\begin{remark}
	Note that in order to employ Theorem~\ref{thm:key-thm}, one can first a-priori fix the desired threshold $\varepsilon$ in~\eqref{eq:metric_lit}. According to the values of $\mathscr{H}$, $\mathscr{L}$, and $T$, one computes the required discretization parameter as $\delta=\frac{\varepsilon}{T\mathscr{H}\mathscr{L}}$. For instance in the case of a uniform quantizer, one can divide each dimension of the set $X$ into intervals of size $\delta/\sqrt{n}$ with $n$ being the dimension of the set.
\end{remark}

\section{Synthesis via Reinforcement Learning}
\label{sec:rl}

\begin{figure*}
  \centering
  \begin{tikzpicture}[every text node part/.style={align=center}]

    \node () at (6.5, 4.8) {Discrete-Time \\Stochastic Control Systems};
    \node[fill=cyan!20,inner sep=5mm,align=center] (CS) at (6.5, 3.5)
    {
      {%
                  $\Sigma=(X,U,\varsigma,f)$}
      };    
    \begin{scope}[yshift=3cm,xshift=-2cm,scale=0.9,every
      node/.style={scale=0.9}, transform shape]
      % Nodes.
      \node[state,initial] (G) {$0$};
      \node[state] (B) [right=2cm of G]{$1$};
     \node[state,accepting] (C) [right=2cm of B] {$2$};
     \node[state] (D) [above=1.5cm of B] {$3$};
      % Transitions.
      \path[->]
      (G) edge node {$\neg \mathtt{b}$} (B)
      (G) edge node {$\mathtt{b}$} (D)
      (B) edge node {$\mathtt{b}$} (D)
      (B) edge node {$\neg \mathtt{b}$} (C)
      (D) edge [loop above] node {$\top$} ()
      (C) edge [loop above] node {$\top$} (C);

    \end{scope}
    \node[fill=cyan!20,inner sep=5mm,align=center] (MJ) at (-0.5, 0)
    {\textsc{Interpreter}};
    \node[single arrow, fill=blue!25] at (4,0.2) {$\delta$-quantized Observations, Reward};
    \node[single arrow, rotate=40, shape border
    rotate=180,fill=blue!25] at (3,2) {~~~~~~~~~~~State~~~~~~~~~~~};
    \node[single arrow, fill=blue!25, rotate=90] at (8,1.75) {Action};
    \node[double arrow, fill=blue!25, rotate=90] at (0,1.5)
    {\phantom{state}};
    \node[inner sep=0pt] (Robot) at (8,0)
    {\includegraphics[height=2cm]{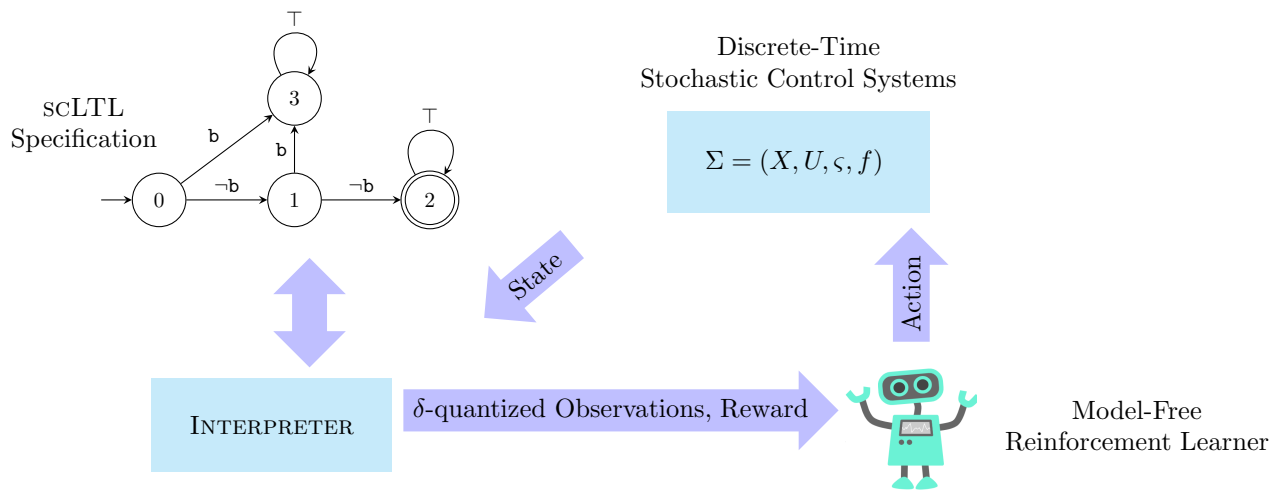}};
    \node (Learner) [right=3cm of Robot] {Model-Free\\ Reinforcement Learner};
    \node (Objective) at (-3,4) {\textsc{scLTL}\\ Specification};
  \end{tikzpicture}
  \vspace{-0.3cm}
  \caption{\small Model-free reinforcement learning is employed
by DFA $\mathcal{A}_\phi$ corresponding to \textsc{scLTL}
    objective $\phi$ to provide scalar rewards by combining DFA $\mathcal{A}_\phi$ and a
$\delta$-quantized observation set of the continuous-space MDP $\Sigma$. In particular,
the $\delta$-quantized observation set of the continuous-space MDP $\Sigma$ is used by an \emph{interpreter} process to
compute a run of $\mathcal{A}_\phi$.
When the run of $\mathcal{A}_\phi$ reaches a final state, the interpreter gives the reinforcement learner a positive
reward and the training episode terminates.
Any converging reinforcement learning
algorithm over such $\delta$-quantized
observation set is guaranteed to maximize the probability of satisfaction of the scLTL
objective $\phi$ and converge to a $2\varepsilon$-optimal strategy over the concrete dt-SCS $\Sigma$
thanks to Theorem~\ref{thm:key-thm}.}
  \label{fig:rl-diagram}
  \vspace{-0.3cm}
\end{figure*}

In this section we sketch how we apply Theorem~\ref{thm:key-thm} to solve
Problem~\ref{problem} when conditional stochastic kernels are unknown.
We begin by detailing the solution of finding optimal policies for scLTL
properties in the case of known MDPs, and then we show how to exploit that to provide a
reinforcement learning-based algorithm to synthesize an optimal policy.

\subsection{Product MDP}
It follows from Theorem~\ref{thm:key-thm} that one can construct a finite
MDP $\widehat \Sigma$ from a continuous-space dt-SCS $\Sigma$ with
known conditional stochastic kernels such that the optimal probability of satisfaction of an
scLTL specification $\phi$ for $T$ steps in $\widehat \Sigma$ is no more than
$2\varepsilon$-worse than the optimal policy in $\Sigma$; see the definition of $\varepsilon$ in Theorem~\ref{thm:key-thm}.
Hence, given a dt-SCS $\Sigma$ with known conditional stochastic kernels, an scLTL property
$\phi$, and a time-horizon $T$, a $2\varepsilon$-optimal policy to satisfy
$\phi$ in $T$ steps is computed using a suitable finite MDP with the corresponding $\delta$ as
the state discretization parameter.
This problem can be solved using the finite-horizon dynamic programming over the
product of $\widehat \Sigma$ and the DFA $\mathcal{A}_\phi$ (cf. Definition~\ref{def:scLTL} and the paragraph afterward) by giving a scalar reward
to all transitions once a final state of $\mathcal{A}_\phi$ is reached.

\begin{definition}[Product MDP]
\label{prod-mdp}
Given a finite MDP $\widehat\Sigma =(\hat X,\hat U, \hat T)$ with initial state
${\hat x_0} \in \hat X$, a labeling function $\Lab: X \rightarrow \alphabeth$ (cf. Subsection \ref{property}),
and  a DFA $\mathcal A_\phi  = (Q, \mathsf{\Sigma}_{\textsf{a}}, \mathsf{t},
q_0, F_{\textsf{a}})$ capturing the scLTL specification $\phi$,  we
define the \emph{product MDP} $\Mm_{\star}$ as a finite MDP $(X_\star, U_\star, T_\star, x_\star, \rho_{\star})$ 
where:
\begin{itemize}
\item $X_\star = \hat X \times Q$ is the set of states;
\item $U_\star = \hat U$ is the set of actions;
\item $T_{\star} : X_\star \times U_\star\times X_\star \to [0,1]$ is the
  probabilistic transition function defined as 
\[
  T_\star((x, q), \nu,({x}',{q}'))
= \begin{cases}
    \hat T(x, \nu, x'), & \text{if } q' = \mathsf{t}(q, \Lab(x)),\\
    0, & \text{otherwise.}
  \end{cases}
\]
  \item $x_\star = (x_0, q_0)$ is the initial state; and
  \item $\rho_\star: X_\star \times U_\star \times X_\star \to \mathbb{N}$ is the reward function
    defined as:
    \[
    \rho_\star((x, q), \nu, (x', q')) =
    \begin{cases}
      1, & \text{ if $q \not \in F$ and $q' \in F$,}\\
      0, &\text{ otherwise.}
    \end{cases}
    \]
\end{itemize}
\end{definition}
Recall that the DFA $A_\phi$ corresponding to an scLTL specification
$\phi$ has the property that there is a unique accepting state and 
all out-going transitions from that state are self-loops.
It follows that total optimal expected reward in the product is equal to the
optimal probability of satisfying the specification.

\begin{proposition}[Product Preserves Probability~\cite{Courco95}]
\label{prop:product-mdp}
  An expected reward-optimal policy in  
  $(X_\star, U_\star, T_\star, x_\star,\\ \rho_{\star})$ along with $A_\phi$
  characterizes an optimal policy in $\widehat\Sigma$ to satisfy $\phi$.
  The optimal expected total reward and an optimal policy can be computed in the
  polynomial time~\cite{papadimitriou1987complexity}.
\end{proposition}

\subsection{Unknown Conditional Stochastic Kernels}

When stochastic kernels are unknown, Theorem~\ref{thm:key-thm} still
provides the correct probabilistic bound given a discretization parameter
$\delta$ if the Lipschitz constant $\mathscr{H}$ is known.    
This observation enables us to employ reinforcement learning algorithms over the
underlying discrete MDP without explicitly constructing the abstraction by simply
restricting observations of the reinforcement learner to the closest
representative point in the set of partitions (cf. Subsection \ref{algo:MC_app}).
We call such an underlying finite MDP a $\widehat\Sigma_{\delta}$ abstraction.

Model-free reinforcement learning can be employed under such observations by
using the DFA $\mathcal{A}_\phi$ to provide scalar rewards as defined in
Definition~\ref{prod-mdp}. 
The observations of the MDP are used by an \emph{interpreter} process to
compute a run of the DFA.
When the DFA reaches a final state, the interpreter gives the reinforcement learner a positive
reward and the training episode terminates. Since the product MDP 
$\Mm_{\star}$
is a finite MDP, from Proposition~\ref{prop:product-mdp}, it follows that any
correct and convergent RL algorithm that maximizes this expected reward is guaranteed to
converge to a policy that maximizes the probability of satisfaction of the scLTL
objective.
From Theorem~\ref{thm:key-thm} it then follows that any converging reinforcement learning
algorithm~\cite{jaakkola1994convergence,borkar2000ode} over such finite
observation set then converges to a $2\varepsilon$-optimal policy over the concrete dt-SCS $\Sigma$
thanks to Theorem~\ref{thm:key-thm}. We summarize the proposed solution in the following theorem.

\begin{theorem}\label{main_theorem}
  Let $\phi$ be an scLTL formula, $\varepsilon>0$, and $\Sigma=(X,U,\varsigma,f)$ be a continuous-space MDP, where $f$ and the distribution of $\varsigma$ are unknown but the Lipschitz constant $\mathscr{H}$ as discussed before is known. For a discretization parameter $\delta$ satisfying $T \delta \mathscr{H}\mathscr{L}\leq\varepsilon$, a convergent model-free
  reinforcement learning algorithm (e.g. Q-learning~\cite{borkar2000ode} or
  TD($\lambda$)~\cite{jaakkola1994convergence}) over $\widehat\Sigma_\delta$ with
  a reward function guided by the DFA $\mathcal{A}_\phi$, converges to a
  $2\varepsilon$-optimal policy over $\Sigma$.
\end{theorem}

\subsection{Reward Shaping: Overcoming Sparse Rewards}
Consider a finite MDP $\widehat\Sigma =(\hat X,\hat U, \hat T)$, a
co-safety automaton
$\mathcal{A}_{\phi}  = (Q, \mathsf{\Sigma}_{\textsf{a}}, \mathsf{t}, q_0, q_F)$, and their
product MDP $\Mm_\star = (X_\star, U_\star, T_\star, x_\star, \rho_{\star})$.
Since the reward function $\rho_\star$ is sparse, it may not be effective in
the reinforcement learning.
For this reason, we introduce a ``shaped'' reward function $\rho_\kappa$
(parameterized by a hyper-parameter $\kappa$) such that for suitable values of
$\kappa$, optimal policies for $\rho_\kappa$ are the same as optimal policies for
$\rho_\star$, but unlike $\rho_\star$ the function $\rho_\kappa$ is dense.

The function $\rho_\kappa$ is defined based on the structure of co-safety automaton $\mathcal{A}_{\phi}$. 
Let $d(q)$ be the minimum distance of the state $q$ to the unique accepting
state $q_F$.
Let $d_{\texttt{max}} = 1 + \max_q \{ d(q) \::\: d(q) < \infty \}$.
If there is no path from $q$ to $q_F$, let $d(q)$ be equal to $d_{\texttt{max}}$.
We define the potential function $P: \mathbb{N} \to \mathbb{R}$ as the following:
\[
P(d) =
\begin{cases}
\kappa \frac{d - d(q_0)}{1 - d_{\texttt{max}}}, &\quad \text{ for $d > 0$},\\
1, &\quad \text{ for $d = 0$},
\end{cases}
\]
where $\kappa$ is a constant hyper-parameter.
Note that the potential of the initial state $P(d(q_0)) = 0$ and the potential
of the final state $P(d(q_F)) = 1$.
Moreover, note that
\begin{equation*}
        P(1) - P(d_{\texttt{max}}) = \kappa.
\end{equation*}
We define the ``shaped'' reward function $\rho_\kappa: \hat X \times \hat U \times \hat
X \to \mathbb{R}$ as the difference between potentials of the destination and of
the target states of transition of the automaton, i.e.,
\[
\rho_\kappa((x, q), \nu, (x', q')) = P(d(q')) - P(d(q)). 
\]
Moreover, notice that for every run $r = (x_0, q_0), \nu_1, (x_1, q_1), $ $\nu_2, \ldots, \nu_n, (x_n,
q_n)$ of $\Mm_\star$, its accumulated reward is simply the potential difference
between the last and the first states, i.e., $P(d(q_n)) - P(d(q_0))$.

\begin{restatable}[Correctness of Reward Shaping]{theorem}{shapingthm}
\label{reward-shaping}
For every product MDP $\Mm_\star =
  (X_\star, U_\star, T_\star, x_\star, \rho_{\star})$, there exists
  $\kappa_\star > 0$ such that for all $\kappa < \kappa_\star$ we have that the set
  of optimal expected reward policies for $\Mm_\star$ is the same as the set of
  optimal expected reward policies for $\Mm_\kappa = (X_\star, U_\star, T_\star,
  x_\star, \rho_\kappa)$.
\end{restatable}
The proof of Theorem~\ref{reward-shaping} is provided in Appendix.
\begin{prnt-cready}
\begin{proof}
  First we note that for the optimality, it is sufficient~\cite{Put94}  to focus on positional
  strategies. 
  Let $\mu_1$ and $\mu_2$ be two positional strategies such that the optimal probability of
  reaching the final state $q_F$ for $\mu_1$ is greater than that for $\mu_2$. We
  write $p_1$ and $p_2$ for these probabilities and $p_1 > p_2$. Notice that these probabilities
  are equal to optimal expected reward with the $\rho_\star$ reward function.

  We denote the expected total reward for policies $\mu_1$ and $\mu_2$ for the
  shaped reward function $\rho_\kappa$ as $s_1$ and $s_2$, respectively.
  These rewards satisfy the following inequalities:
  \begin{eqnarray*}
    s_1 \!\!&\geq&\!\! p_1 (P(0) {-} P(d(q_0))) + (1{-}p_1) (P(d_{\texttt{max}}){-}P(d(q_0)),\\
    s_2 \!\!&\leq&\!\! p_2 (P(0) {-} P(d(q_0)))  + (1{-}p_2) (P(1){-}P(d(q_0))).
  \end{eqnarray*}
  It can be verified that if $\kappa < p_1 - p_2$, then $s_1 > s_2$.
  Therefore, if $\mu_1$ is an optimal positional strategy, and $\mu_2$ is one of the next best
  positional strategies, choosing $\kappa_* < p_1 - p_2$ guarantees that an optimal strategy in
  $\Mm_\kappa$ is also optimal for $\Mm_\star$.
\end{proof}
\end{prnt-cready}

Theorem~\ref{reward-shaping} demonstrates one way to shape rewards such that the
optimal policy remains unaffected while making the rewards less sparse.
Along similar lines, one can construct a variety of potential functions and
corresponding shaped rewards with similar correctness properties.
Of course, the reward shaping schema presented here is no silver bullet: we
expect the performance of different potential functions to be incomparable along
a carefully chosen ensemble of MDPs.
Since rewards are shaped without any knowledge of the underlying MDP, there
may be MDPs where un-shaped rewards may work as well or even 
better than a given shaped reward. We envisage that the ability to combine several competing
ways to shape reward may work better in practice.
While sparse rewards may be sufficient for simpler learning tasks, we
demonstrate that shaped rewards such as the one provided here are crucial for
larger case studies such as the BMW case-study reported in the next section.

\section{Case Studies}

Before illustrating our results via some experiments, we elaborate on
the dimension dependency in our proposed RL techniques compared to the
abstraction-based ones. Assuming a uniform quantizer, the finite MDP
constructed in Subsection~\ref{algo:MC_app} is a matrix with a
dimension of $(n_x\times n_u)\times n_x$, where $n_u$ is the
cardinality of the finite input set $U$. Computing this matrix is one
of the bottlenecks in abstraction-based approaches since an
$n$-dimensional integration has to be done numerically for each
entries of this matrix.  Moreover, $n_x$ (i.e., the cardinality of the finite
state set) grows exponentially with the dimension $n$. Once this
matrix is computed, it is employed for the dynamic programming on a
vector of the size $(n_x\times n_u)$.  This is a second bottleneck of the
process.  On the other hand, by employing the proposed RL approach,
the curse of dimensionality reduces to only \emph{learning} the vectors
of size $(n_x\times n_u)$ without having to compute the full
matrix. Moreover, the abstraction-based techniques need to precisely
know the probabilistic evolution of states in models, whereas
in this work we only need to know the Lipschitz constant
$\mathscr{H}$.

Concerning the trade-off between iteration count, discretization size,
and performance, we should mention that by decreasing the
discretization parameter, the closeness error in
Theorem~\ref{thm:key-thm} is reduced. On the other hand, one needs
more training episodes as the size of the problem increases. Note that
in our proposed setting, we do not need to compute transition
probabilities $\hat T$ for finite MDPs $\widehat\Sigma$, since we
directly learn the value functions using RL.

To demonstrate the effectiveness of the proposed results, we first apply our proposed approaches
to two physical benchmarks including regulation of room
temperature and control of road traffic.
We then apply our algorithms to a nonlinear model of a BMW $320$i car by
synthesizing a controller enforcing a reach-while-avoid specification. The
first two case studies are intentionally chosen to be small such that we can
compare (cf. Table \ref{tab:results}) probabilities of satisfaction $p_r$
estimated by RL with the optimal probabilities $p_*$ computed using the dynamic
programming when the exact dynamics are known. 

\subsection{Room Temperature Control}

Here, we apply our results to the temperature regulation of a room equipped with a heater. The model of this case study is adapted from \cite{meyer} by including
stochasticity in the model as an additive noise.
The evolution of the temperature can be described by the following dt-SCS:
\begin{align}\notag
\Sigma:{x}(k+1)=&(1-2\eta-\beta-\gamma\nu(k)){x}(k)+\gamma T_{h}\nu(k)+ \beta T_{e}+0.3162\varsigma(k),
\end{align}
where $\eta = 0$, $\beta = 0.022$, and $\gamma = 0.05$ are conduction factors respectively between this room and the other rooms in a network,
between the external environment and the room,
and between the heater and the room.
Moreover, $x(k)$ and $\nu(k)$ are taking values in $[19,21]$ and a finite input set~$\{0.03 , 0.09 , 0.15, 0.21 , 0.27 , 0.33 , 0.39,\\ 0.45 , 0.51 , 0.57\},$ respectively. The parameter $T_{e}=-1\,^\circ C$ is the outside temperature, and $T_h=50\,^\circ C$ is the heater temperature.

The main goal is to synthesize a controller for $\Sigma$ using our main results in Theorem \ref{main_theorem} such that the controller maintains the temperature of the room in the safe set $[19,21]$.

\subsection {Road Traffic Control}

\begin{figure}[ht]
	\begin{center}
		\includegraphics[width=6.5cm]{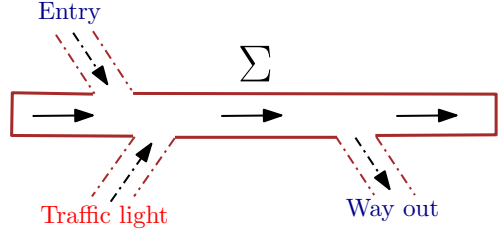}
		\caption{Model of a road traffic control with the length of $500$ meters, $1$ way out, and $2$ entries, one of which is controlled by a traffic light.} 
		\label{Fig2}
	\end{center}
	\vspace{-0.5cm}
\end{figure}

We also apply our results to a road traffic control containing a cell with $2$ entries and $1$ way out, as
schematically depicted in Figure~\ref{Fig2}. The model of this case
study is taken from~\cite{le2013mode}; stochasticity is included in
the model as the additive noise.
One of the entries of the cell is controlled by a traffic light, denoted by $\nu
= \{0,1\}$, that enables (green light) or not (red light) the vehicles to
pass. In this model, the length of a cell is $0.5$ kilometers ([km]), and the
flow speed of the vehicles is $100$ kilometers per hour ([km/h]). Moreover, 
during the time interval $\tau = 6.48$ seconds, it is assumed that $6$ vehicles pass the entry controlled by
the traffic light, $3$ vehicles go into the entry of the cell, and one quarter of vehicles goes out on the exit of the cell (the ratio denoted by $q$). We want to observe the density of traffic $x$. 
The model of the system $\Sigma$ is described by:
\begin{align}\notag
\Sigma:x(k+1) = (1-\frac{\tau v}{l}-q)x(k)+ 6\nu(k)+ 1.9494\varsigma(k)+3,
\end{align}
where $l$ and $v$ are the length of the cell and the flow speed of vehicles, respectively.
We synthesize a controller for $\Sigma$ using our main results in Theorem \ref{main_theorem} such that the density of the traffic is lower than $20$ vehicles.

\subsection{Experiments}\label{example}

\begin{table*}
	\centering
	\renewcommand{\arraystretch}{1.1}
	\caption{Q-Learning Results for Room Temperature and Road Traffic Examples.}
	\vspace{-0.3cm}
	\label{tab:results}
	\begin{tabular}{c*{2}{|ccccc}}
		& \multicolumn{5}{|c}{\texttt{Room}} &
		\multicolumn{5}{|c}{\texttt{Traffic}} \\
		$\delta$ & $p_r$ & $p_*$ &  $\varepsilon$ & $p_l$& $p_h$ &$p_r$ & $p_*$ & $\varepsilon$ & $p_l$ & $p_h$ \\\hline
		$0.01$ & 0.9698 & 0.9753 & 0.2468 & 0.7285 & 1.0 & 0.9856 & 0.9995 & 0.0160 & 0.9835 & 1.0 \\
		$0.02$ & 0.9745 & 0.9753 & 0.4936 & 0.4817 & 1.0 & 0.9975 & 0.9995 & 0.0319 & 0.9676 & 1.0 \\
		$0.05$ & 0.9543 & 0.9753 & 1.2339 & 0.0000 & 1.0 & 0.9993 & 0.9995 & 0.0798 & 0.9197 & 1.0 \\
		$0.1$  & 0.9779 & 0.9754 & 2.4678 & 0.0000 & 1.0 & 0.9999 & 0.9995 & 0.1596 & 0.8399 & 1.0 \\
		$0.2$  & 0.9732 & 0.9743 & 4.9357 & 0.0000 & 1.0 & 0.9999 & 0.9995 & 0.3193 & 0.6802 & 1.0 \\
	\end{tabular}
	\vspace{-0.3cm}
\end{table*}

\begin{figure*}
	\centering
	\includegraphics[height=0.37\textwidth,width=0.49\textwidth]{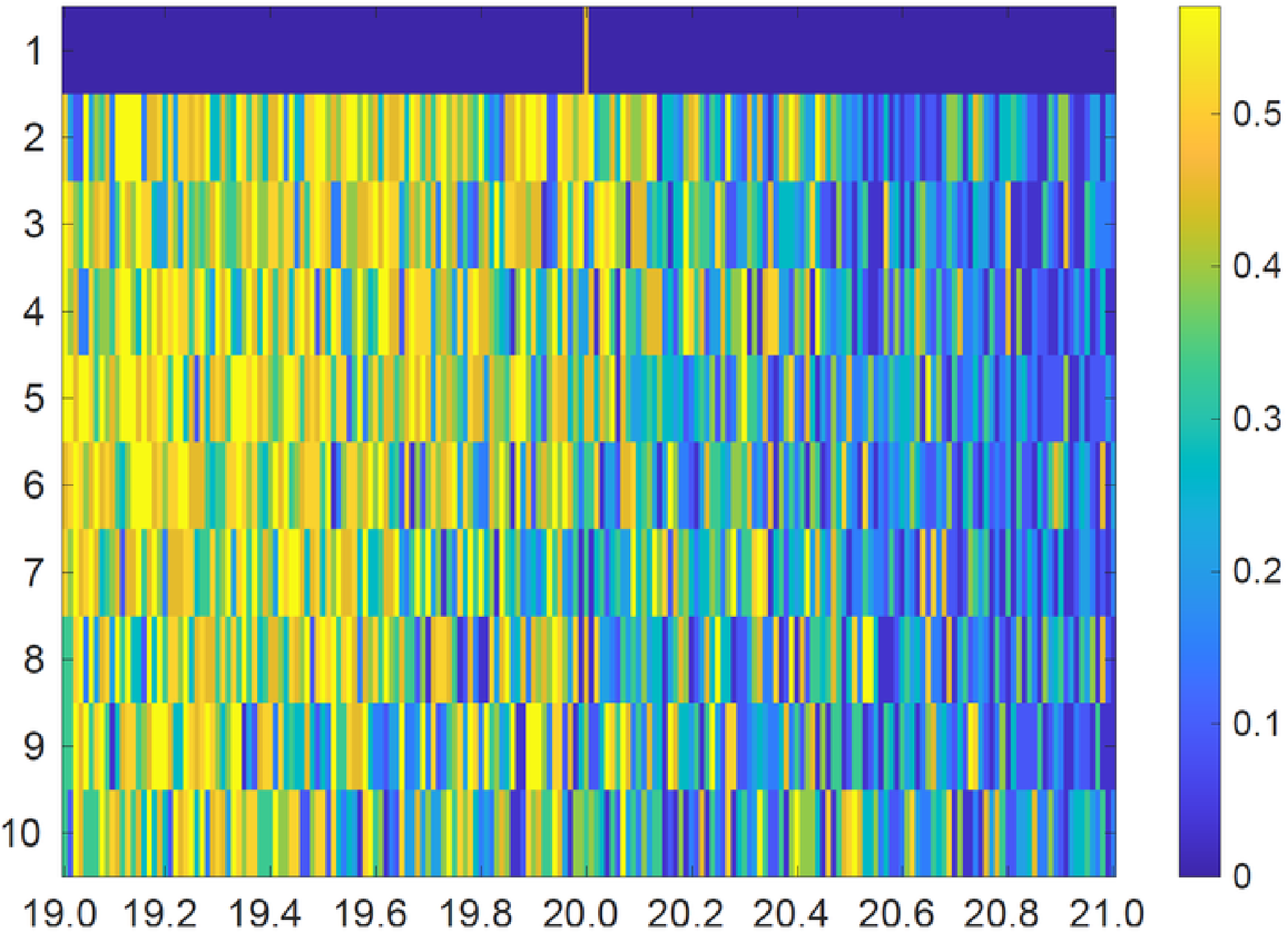}\hspace{0.1cm}
	\includegraphics[height=0.37\textwidth,width=0.49\textwidth]{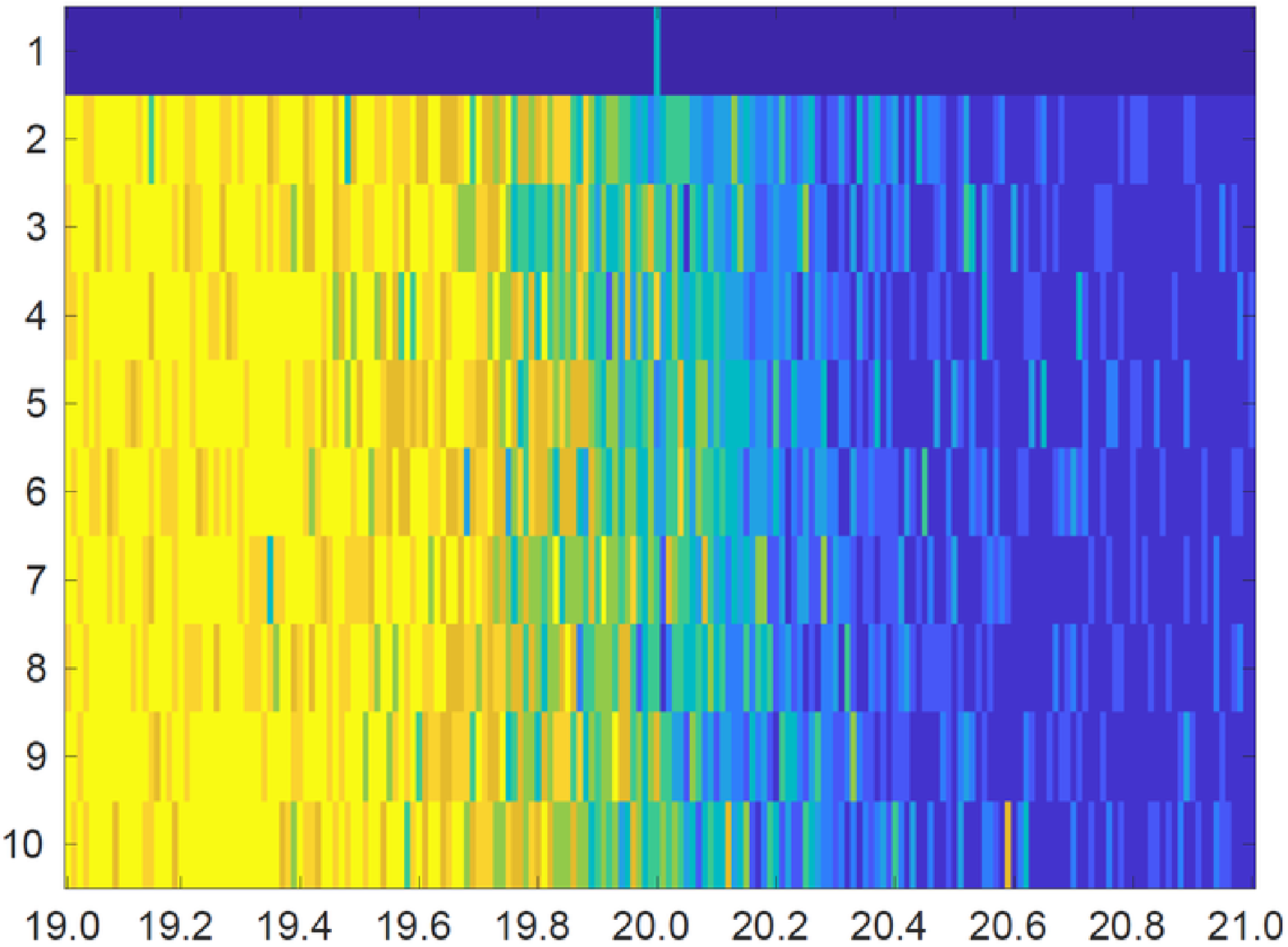}
	\caption{Room temperature control: A heatmap visualization of
		strategies learned via Reinforcement Learning after $10^5$
		episodes (left) and after $8 \cdot 10^6$ episodes (right).
		The $X$ axis represents the room temperature in
		${}^{\circ}\!\mathsf{C}$, while the $Y$ axis represents time
		steps $1 \leq k \leq 10$. The action suggested by the
		strategy is in the finite input set
		$\{0.03, 0.09 , 0.15, 0.21 , 0.27 , 0.33 , 0.39 , 0.45 ,
		0.51 , 0.57\}$ and is color-coded according to the map shown
		in the middle: Bright yellow and deep blue represent maximum
		and minimum heat.  In the first step, strategies are
		only defined for the initial state; this causes the blue
		bands at the top.}
	\label{fig:room-strategy}
	\vspace{-0.5cm}
\end{figure*}

Table~\ref{tab:results} shows a comparison of Q-learning to the
computed optimal probabilities for the room temperature and road traffic examples.
For each model, five different discretization steps ($\delta$) are
considered and for each value of $\delta$ probabilities of
satisfaction of the safety objectives are reported in the columns
labeled $p_r$.  These probabilities are $Q$-values of the initial
state of the finite-state MDP for the policy computed by $Q$-learning
after $10^6$ episodes.  The objective is to keep the system safe for
at least $10$ steps. For the comparison, the optimal probability $p_*$ for
a time-dependent policy is reported assuming that we know the exact
dynamics for these two examples.  Note that we compute $p_*$ using
the dynamic programming over constructed finite MDPs as proposed in
Subsection~\ref{algo:MC_app}. The optimal probability $p_*$ reported in
Table~\ref{tab:results} corresponds to the same initial condition that
is utilized in the learning process. The optimal probability for the
original \emph{continuous-space} MDP is always within an interval $[p_l,p_h]$
centered at $p_*$ and with a radius $\varepsilon$ as reported in
Table~\ref{tab:results}. One can readily see from
Table~\ref{tab:results} that as the discretization parameter $\delta$
decreases, the size of this interval shrinks, which implies that the
optimal probability for the original \emph{continuous-space} MDP
converges to $p_*$.  While finer abstractions give better
theoretical guarantees, for a fixed number of episodes it is easier to
learn good strategies for coarser abstractions.  This is reflected in
Table~\ref{tab:results}, where the values of $p_r$ do not necessarily
get better with smaller values of $\delta$.  However, by increasing
the number of episodes, strategies converge toward the optimal
one, as illustrated in Figure~\ref{fig:room-strategy}, which
visualizes room temperature control strategies computed by
$Q$-learning after different numbers of episodes.  Note that in
Table~\ref{tab:results}, the error bound $\varepsilon$ exceeds
one for $\delta \ge 0.05$ in the room temperate control example,
which is not a useful probability bound for the
\emph{continuous-space} MDP. However, we prefer to report the corresponding
values of $p_r$ and $p_*$ so that they can still be compared.

\subsection{7-Dimensional Autonomous Vehicle}

The previous case-studies are representative of what can be solved by
discretization and tabular methods like Q-learning.
Relaxing those constraints, we were able to apply deep deterministic policy
gradient (DDPG)~\cite{Lillicrap15} to a $7$-dimensional \emph{nonlinear} model
of a BMW $320$i car~\cite{Althof18} to synthesize a
reach-while-avoid controller.
Though convergence guarantees are not available for DDPG and most RL
algorithms with nonlinear function approximations,
breakthroughs in this direction (e.g., SBEED in~\cite{dai2017sbeed}) will expand
the applicability of our results to more complex safety-critical applications.

The model of this case study is borrowed from~\cite[Section
5.1]{Althof18} by discretizing the dynamics in time and including
a stochasticity inside the dynamics as additive noises.
\begin{prnt-arxiv}
	The dynamics of the vehicle are given in Appendix.
\end{prnt-arxiv}
We are interested in an autonomous operation of the vehicle on a
highway. Consider a situation on a two-lane highway when an accident suddenly
happens on the same lane on which our vehicle is traveling. The vehicle's
controller should find a safe maneuver to avoid the crash with the
next-appearing obstacle. 

\begin{figure*}
	\centering
	\includegraphics[width=0.95\textwidth]{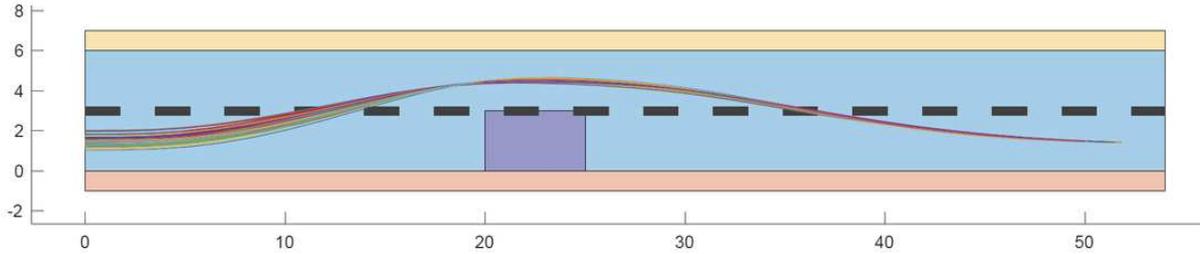}
	\caption{Trajectories of $100$ simulations of the RL-synthesized controller
		for a $7$-dimensional model of a BMW $320i$ car trained using  DDPG. The road is
		$6$ meter wide and $50$ meter long, and the length of the car is $4.508$
		meters and its width is $1.610$ meters.
	}
	\label{fig:trainedBeemers}
	\vspace{-0.3cm}
\end{figure*}

Figure~\ref{fig:trainedBeemers} shows the simulation from $100$ samples with varying
initial positions and initial heading velocities ($16$--$18$ m/s) for the
learned controller.
We employed potential-based reward shaping to speed-up learning in this case study
from 10K episodes (no success) to under 5K episodes (for a convincing learning,
see Figure~\ref{fig:trainedBeemers}).

\section{Conclusion}

We studied the problem of finding policies for systems that can be
modeled as continuous-space MDPs but with unknown dynamics.
The goal of the policy is to maximize the probability that the system satisfies
a complex property expressed as a fragment of linear temporal logic
formulae.
Our approach replaces the unknown system with a finite MDP without explicitly
constructing it.
Since  transition probabilities of the finite MDP are unknown, we utilize
the reinforcement learning (RL) to find a policy and apply it to the original
continuous-space MDP.
We show that any converging reinforcement learning
algorithm~\cite{jaakkola1994convergence,borkar2000ode} over such finite
observation MDP converges to a $2\varepsilon$-optimal strategy over the concrete 
continuous-space MDP with unknown dynamics (only an upper bound on the Lipschitz
constant is known), where $\varepsilon$ is defined a-priori and can be
controlled.
We applied our approach to multiple case studies. The results are promising and
demonstrate that by employing an automata-theoretic reward shaping, the learning
algorithm enlarges the class of systems over which we can
perform the formal synthesis.

\bibliographystyle{alpha}
\bibliography{biblio}

\section{Appendix}

\begin{IEEEproof}\textbf{(Theorem~\ref{reward-shaping})}
First we note that for the optimality, it is sufficient~\cite{Put94}  to focus on positional
strategies. 
Let $\mu_1$ and $\mu_2$ be two positional strategies such that the optimal probability of
reaching the final state $q_F$ for $\mu_1$ is greater than that for $\mu_2$. We
write $p_1$ and $p_2$ for these probabilities and $p_1 > p_2$. Notice that these probabilities
are equal to the optimal expected reward with the $\rho_\star$ reward function.

We denote the expected total reward for policies $\mu_1$ and $\mu_2$ for the
shaped reward function $\rho_\kappa$ as $s_1$ and $s_2$, respectively.
These rewards satisfy the following inequalities:
\begin{eqnarray*}
	s_1 \!\!&\geq&\!\! p_1 (P(0) {-} P(d(q_0))) + (1{-}p_1) (P(d_{\texttt{max}}){-}P(d(q_0)),\\
	s_2 \!\!&\leq&\!\! p_2 (P(0) {-} P(d(q_0)))  + (1{-}p_2) (P(1){-}P(d(q_0))).
\end{eqnarray*}

Now consider:

\begin{eqnarray*}
	s_1 - s_2 \!\!\!\!&\!\!\! \geq \!\!\!&\!\!\!\! \big(p_1 (P(0) {-} P(d(q_0))) + (1{-}p_1)(P(d_{\texttt{max}}){-}P(d(q_0))\big) - \big(p_2 (P(0) {-} P(d(q_0)))  + (1{-}p_2) (P(1){-}P(d(q_0)))\big).\\
	& = & \big(p_1 (1 {-} P(d(q_0))) + (1{-}p_1)(P(d_{\texttt{max}}){-}P(d(q_0))\big) - \big(p_2 (1 {-} P(d(q_0)))  + (1{-}p_2) (P(1){-}P(d(q_0)))\big)\\
	& = & \big(p_1 + (1{-}p_1) P(d_{\texttt{max}}) - P(d(q_0))\big) - \big(p_2 + (1{-}p_2) P(1) - P(d(q_0)))\big)\\
	& = & \big(p_1 + (1{-}p_1) P(d_{\texttt{max}}) \big) - \big(p_2 + (1{-}p_2) P(1)\big)\\
	& = & \big(p_1 + (1{-}p_2) P(d_{\texttt{max}}) - (p_1{-}p_2) P(d_{\texttt{max}})
	\big) - \big(p_2 + (1{-}p_2) P(1) \big)\\
	& = & (p_1 - p_2) + (1{-}p_2) (P(d_{\texttt{max}}) - P(1))  - (p_1{-}p_2) P(d_{\texttt{max}})\\
	& = & (p_1 - p_2) - (1{-}p_2) \kappa - (p_1{-}p_2) P(d_{\texttt{max}})\\
	& = & (p_1 - p_2) - \kappa \big((1{-}p_2) + (p_1{-}p_2) P(d_{\texttt{max}})\big)\\
	& \geq & (p_1 - p_2) - \kappa~\text{(since $p_2 \geq 0$, $p_1 - p_2 > 0$, and
		$P(d_{\texttt{max}}) \leq 0$}).\\
\end{eqnarray*}
It can be verified that if $\kappa < p_1 - p_2$ then $s_1 > s_2$.
Therefore, if $\mu_1$ is an optimal positional strategy, and $\mu_2$ is one of the next best
positional strategies, choosing $\kappa_* < p_1 - p_2$ guarantees that an optimal strategy in
$\Mm_\kappa$ is also optimal for $\Mm_\star$.
\end{IEEEproof}

{\bf Dynamics of 7-Dimensional Autonomous Vehicle.} For $\vert x_4(k) \vert < 0.1$:
\begin{align}\notag
x_i(k+1) &= x_i(k) + \tau a_i + 0.5\varsigma_i(k),  i \in \{1,\dots, 7\}\backslash\{3,4\}, \\\notag
x_3(k+1) &= x_3(k) + \tau \text{Sat}_1(\nu_1) + 0.5\varsigma_3(k),\\\notag
x_4(k+1) &= x_4(k) + \tau \text{Sat}_2(\nu_2) + 0.5\varsigma_4(k),
\end{align}
and for $\vert x_4(k) \vert \ge 0.1$:
\begin{align}\notag
x_i(k+1) &= x_i(k) + \tau b_i + 0.5\varsigma_i(k), i \in \{1,\dots, 7\}\backslash\{3,4\}, \\\notag
x_3(k+1) &= x_3(k) + \tau \text{Sat}_1(\nu_1) + 0.5\varsigma_3(k),\\\notag
x_4(k+1) &= x_4(k) + \tau \text{Sat}_2(\nu_2) + 0.5\varsigma_4(k),
\end{align}			
where,
\begin{align}\notag
a_1 &=\! x_4\text{cos}(x_5(k)), a_2 = x_4\text{sin}(x_5(k)), a_5 = \frac{x_4}{l_{wb}} \text{tan}(x_3(k)),			
a_6 = \frac{\nu_2(k)}{l_{wb}} \text{tan}(x_3(k)) + \frac{x_4}{l_{wb}\text{cos}^2(x_3(k)\!)} \nu_1(k),\\\notag
\quad a_7 &= 0, b_1 = x_4(k)\text{cos}(x_5(k)+x_7(k)), b_2 =x_4(k)\text{sin}(x_5(k)+x_7(k)\!), b_5 = x_6(k), \\\notag
b_6 &= \frac{\mu m}{I_z(l_r\!+\!l_f)}(l_fC_{S,f}(gl_r\!-\!\nu_2(k)h_{cg})x_3(k)+(l_rC_{S,r}(gl_f\!+\!\nu_2(k)h_{cg})\!-\!l_fC_{S,f}(gl_r\!-\!\nu_2(k)h_{cg})\!)x_7(k)\\\notag
&\quad-(l_f^2C_{S,f}(gl_r\!-\!\nu_2(k)h_{cg})\!+\!l_r^2C_{S,r}(gl_f\!+\!\nu_2(k)h_{cg})\!)\frac{x_6(k)}{x_4(k)}),\\\notag
b_7 &=\! \frac{\mu}{x_4(k)(l_r\!+\!l_f)}(C_{S,f}(gl_r\!-\!\nu_2(k)h_{cg})x_3(k)+(C_{S,r}(gl_f\!+\!\nu_2(k)h_{cg})+C_{S,f}(gl_r\!-\!\nu_2(k)h_{cg})\!)x_7(k)\\\notag
&\quad-(l_fC_{S,f}(gl_r\!-\!\nu_2(k)h_{cg})\!-\!l_rC_{S,r}(gl_f\!+\!\nu_2(k)h_{cg})\!)\frac{x_6(k)}{x_4(k)})-x_6(k).
\end{align}
Moreover, $\text{Sat}_1(\cdot)$ and $\text{Sat}_2(\cdot)$ are input saturation functions introduced by \cite[Section 5.1]{Althof18},
$x_1$ and $x_2$ are the position coordinates, 
$x_3$ is the steering angle, 
$x_4$ is the heading velocity, 
$x_5$ is the yaw angle, 
$x_6$ is the yaw rate, and 
$x_7$ is the slip angle. 
Variables $\nu_1$ and $\nu_2$ are inputs and they control the steering angle and heading velocity, respectively. 

The model takes into account the tire slip making it a good candidate for studies that consider planning of evasive maneuvers that are very close to physical limits. 
We consider an update period $\tau = 0.001$ [s] and the following parameters for a BMW $320$i car: 
$l_{wb} = 2.5789$ as the wheelbase [m],
$m = 1093.3$ [kg] as the total mass of the vehicle,
$\mu = 1.0489$ as the friction coefficient, 
$l_f = 1.156$ [m] as the distance from the front axle to center of gravity (CoG), 
$l_r = 1.422$ [m] as the distance from the rear axle to CoG, 
$h_{cg} = 0.574$ [m] as the hight of CoG, 
$I_z = 1791.6$ [kg m$^2$] as the moment of inertia for entire mass around $z$ axis, 
$C_{S,f} = 20.89$ [$1$/rad] as the front cornering stiffness coefficient, and 
$C_{S,r} = 20.89$ [$1$/rad] as the rear
cornering stiffness coefficient.

We consider a bounded set $X := [0,84] {\times}[0,6]{\times}$ $[-0.18,0.18]{\times}[12,21]{\times}[-0.5,0.5]{\times}[-0.8,0.8]{\times}[-0.1,
0.1]$, and a quantized input
set $U := [-0.4,0.4]{\times}[-4,4]$ with a fine quantization parameter.

\end{document}